\documentclass[aps,prd,superscriptaddress,preprintnumbers]{revtex4}

\usepackage{graphicx}
\newcommand*{\ie}{\textit{i.e.},\ }
\newcommand*{\etal}{\textit{et al.\ }}

\newcommand*{\gevc}{\,\textrm{GeV/c}}

\newcommand*{\tev}{\,\textrm{TeV}}
\newcommand*{\pt}{$p_T$}

\newcommand*{\BBR}{``beam-beam remnants"}

\newcommand*{\pthard}{$p_T({\rm hard})$}

\newcommand*{\ycut}{$|y|\!<\!1$}
\newcommand*{\hardcut}{$p_T\!(\textrm{hard})>3{\rm\,GeV/c}$}

\newcommand*{\etaphi}{$\eta$-$\phi$}
\newcommand*{\bbar}{$\bar b$}
\newcommand*{\bpair}{$b\bar b$}

\newcommand*{\qsq}{$Q^2$}
\newcommand*{\delphi}{$|\Delta\phi|$}
\begin{document}

\preprint{UFIFT-HEP-01-25}
%
\title{The Sources of b-Quarks at the Tevatron and their Correlations}

\author{R. D. Field}
\email[]{rfield@phys.ufl.edu}
\homepage[]{http://www.phys.ufl.edu/~rfield/}
\affiliation{
Department of Physics, University of Florida,
Gainesville, Florida, 32611, USA}
\date{\today}

\begin{abstract}
The leading-log order QCD hard scattering Monte-Carlo models of HERWIG, ISAJET, and PYTHIA are used to 
study the sources of $b$-quarks at the Tevatron.  The reactions responsible for producing $b$ and \bbar\ quarks are separated 
into three categories; flavor creation, flavor excitation, and parton-shower/fragmentation. Flavor creation 
corresponds to the production of a \bpair\ pair by gluon fusion or by annihilation of light quarks, while flavor 
excitation corresponds to a $b$ or  \bbar\ quark being knocked out of the initial-state by a gluon or a light quark or 
antiquark.  The third source occurs when a \bpair\ pair is produced within a parton-shower or during the fragmentation 
process of a gluon or a light quark or antiquark (includes gluon splitting).  The QCD Monte-Carlo models indicate 
that all three sources of $b$-quarks are important at the Tevatron and when combined they qualitatively describe the
inclusive cross-section data. Correlations between the $b$ and  \bbar\ quark are 
very different for the three sources and can be used to isolate the individual contributions.
\end{abstract}

\maketitle

\section{Introduction}

It is important to have good leading order (or leading-log order) estimates of hadron-hadron collider 
observables.  Of course, precise comparisons with data require beyond leading order calculations.  If 
the leading order estimates are within a factor of two of the data, higher order calculations might be 
expected to improve the agreement.  On the other hand, if the leading order estimates are off by more 
than about a factor of two of the data, one cannot expect higher order calculations to improve the 
situation.  In this case, even if the higher order corrections were large enough to bring agreement, 
one could not trust a perturbative series in which the second term is greater than the first.  If a 
leading order estimate is off by more than a factor of two, it usually means that one has overlooked 
some important physics.  For this reason good leading-log order estimates are important.  

\begin{figure}[htbp]
\includegraphics[scale=0.6]{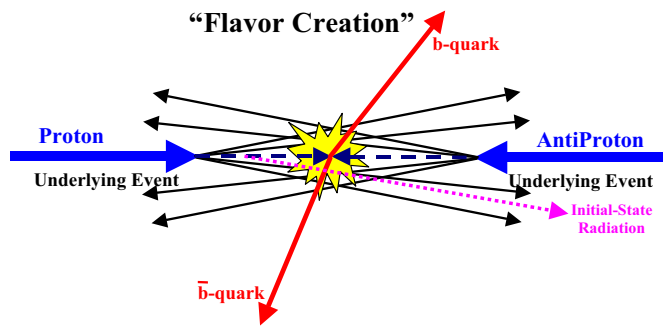}
\caption{Illustration of a proton-antiproton collision in which the QCD hard $2$-to-$2$ reaction results in 
the creation of a \bpair\ pair via the subprocess $q+\bar q\to b+\bar b$ or $g+g\to b+\bar b$.  The event also
contains possible initial and final-state radiation and particles that result from the break-up of the initial 
proton and antiproton (\ie \BBR).}
\label{prd_fig1}
\end{figure}

\begin{figure}[htbp]
\includegraphics[scale=0.5]{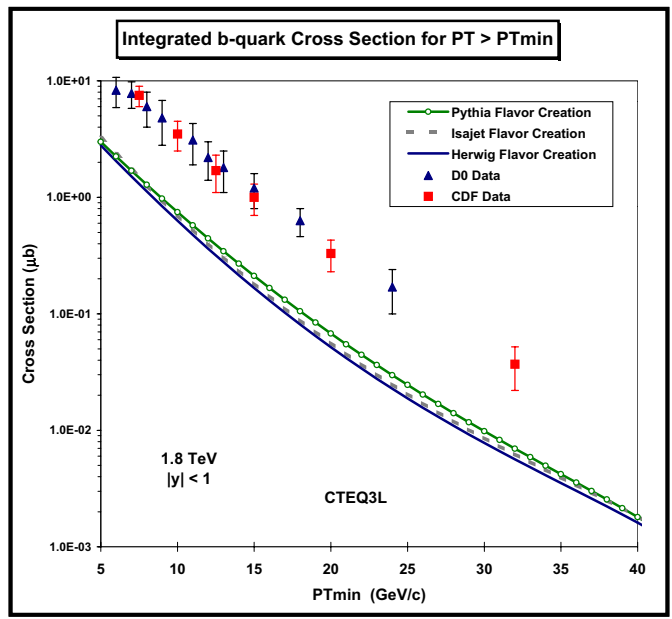}
\caption{Data from CDF and D0 \cite{CDF1,CDF2,D0,wester} for the integrated inclusive $b$-quark total cross section 
($p_T>p_T({\rm min})$, \ycut) for proton-antiproton collisions at $1.8\tev$ compared with the 
QCD Monte-Carlo model predictions of HERWIG 5.9, PYTHIA 6.115, and ISAJET 7.32 for the ``flavor creation" 
subprocesses illustrated in FIG.~\ref{prd_fig1}.  All three Monte-Carlo models were generated
using the parton distribution functions CTEQ3L and \hardcut.}
\label{prd_fig2}
\end{figure}

In this analysis the leading-log order QCD hard scattering Monte-Carlo models of HERWIG \cite{herwig}, 
ISAJET \cite{isajet}, and PYTHIA \cite{pythia} are used to study the sources of $b$-quarks at the Tevatron.  
The reactions responsible of producing $b$-quarks are separated into three categories; flavor 
creation, flavor excitation, and shower/fragmentation.  Flavor creation corresponds to the 
production of a \bpair\ pair by gluon fusion or by the annihilation of light quarks via the 
two $2$-to-$2$ parton subprocesses, $q+\bar q\to b+\bar b$, and $g+g\to b+\bar b$, 
and is illustrated in FIG.~\ref{prd_fig1}.  The data from CDF and D0 \cite{CDF1,CDF2,D0,wester} for the 
integrated inclusive $b$-quark cross section for \ycut\ at $1.8\tev$ are compared with the QCD 
Monte-Carlo model predictions for flavor creation in FIG.~\ref{prd_fig2}, where $y$ is the rapidity of 
the $b$-quark.  Here the parton distribution functions CTEQ3L have been used for all three Monte-Carlo 
models and, as is well know, the leading 
order predictions are roughly a factor of four below the data.  The leading order estimates of the 
flavor creation contribution to $b$-quark production at the Tevatron are so far below the data that 
higher order corrections (\textit{even though they may be important}) cannot be the whole story.

\begin{figure}[htbp]
\includegraphics[scale=0.6]{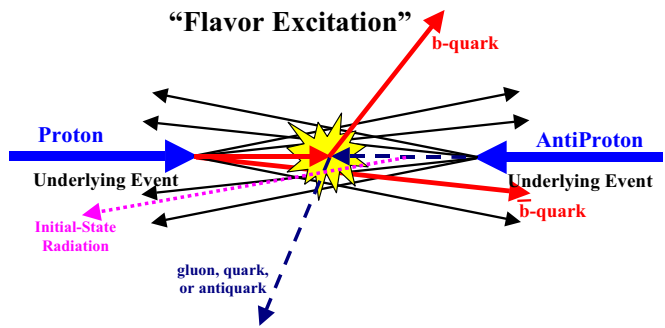}
\caption{Illustration of a proton-antiproton collision in which the QCD hard $2$-to-$2$ reaction corresponds 
to the scattering of a $b$-quark out of the initial-state into the final-state by a gluon or a light 
quark or light antiquark via the subprocess  $g+b\to g+b$, $q+b\to q+b$, or $\bar q+b\to \bar q+b$.  
These subprocesses together with the corresponding \bbar\ terms are referred to as ``flavor excitation".
The event also contains possible initial and final-state radiation and particles that result from the 
break-up of the initial proton and antiproton (\ie \BBR).}
\label{prd_fig3}
\end{figure}

An additional source of $b$-quarks at the Tevatron comes from the scattering of a $b$ or \bbar\ quark out of 
the initial-state into the final-state by a gluon or by a light quark or antiquark via the subprocesses;
$g+b\to g+b$, $g+\bar b\to g+\bar b$, $q+b\to q+b$, $q+\bar b\to q+\bar b$, 
$\bar q+b\to \bar q+b$, and $\bar q+\bar b\to \bar q+\bar b$.
This is referred to as ``flavor excitation" and is illustrated in FIG.~\ref{prd_fig3}.  Flavor excitation 
is, of course, very sensitive to the number of $b$ and \bbar\ quarks within the proton (\ie the structure functions).  
The $b$ and \bbar\ quarks are generated through the \qsq\ evolution of the structure functions.  Even with 
no ``intrinsic"  \bpair\ pairs within the proton, at high \qsq\ \bpair\ pairs are produced by gluons 
and populate the proton ``sea".  The number of \bpair\ pairs within the proton is related, through the \qsq\ evolution, 
to the gluon distribution within the proton.  None of the structure functions considered in this analysis 
include ``intrinsic"  \bpair\ pairs within the proton.  The \bpair\ pair content within the proton is generated 
entirely through the \qsq\ evolution of the structure functions.

\begin{figure}[htbp]
\includegraphics[scale=0.6]{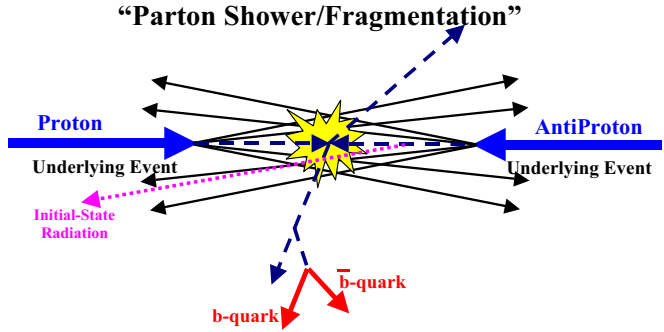}
\caption{Illustration of a proton-antiproton collision in which a \bpair\ pair is created 
within a parton-shower or during the fragmentation process of a gluon or a light quark or antiquark.  
Here the QCD hard $2$-to-$2$ subprocess involves only gluons and light quarks and antiquarks 
(no heavy quarks in the $2$-to-$2$ hard scattering subprocess).  The event also
contains possible initial and final-state radiation and particles that result from the break-up of the initial 
proton and antiproton (\ie \BBR).}
\label{prd_fig4}
\end{figure}

Another source of $b$-quarks at the Tevatron comes from reactions which have a \bpair\ in the 
final-state but with only gluons and light quarks and light antiquarks participating in the $2$-to-$2$ hard 
parton scattering subprocess (\ie no heavy quarks in the $2$-to-$2$ hard scattering subprocess).  This is 
referred to as ``shower/fragmentation" and is illustrated in FIG.~\ref{prd_fig4}.  Here the subprocesses 
are all QCD $2$-to-$2$ gluon, light quark, and light antiquark subprocesses. The ``shower/fragmentation"
contribution comes from \bpair\ pairs produced within parton-showers or during the fragmentation process.  
This category includes the ``gluon splitting" subprocess, $g+g\to g+(g\to b\bar b)$, as modeled by the 
QCD leading-log Monte-Carlo models.

Section II discusses methods of generating the three sources of $b$-quarks; flavor creation, 
flavor excitation, and shower/fragmentation.  In Section III the leading-log QCD Monte-Carlo model predictions 
are compared with data on the $b$-quark inclusive cross section at the Tevatron. Correlations between the $b$ and  
\bbar\ quark are very different for the three sources and in Section IV we study ways to isolate the 
individual contributions using  \bpair\ correlations.  Section V is reserved for summary and conclusions.

\begin{figure}[htbp]
\includegraphics[scale=0.5]{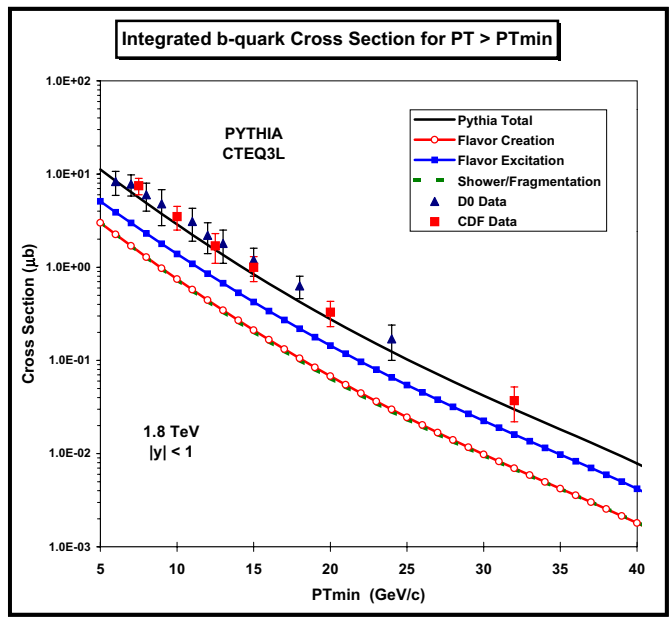}
\caption{Data on the integrated inclusive $b$-quark total cross section  
($p_T>p_T({\rm min})$, \ycut) for proton-antiproton collisions at $1.8\tev$ compared with the 
QCD Monte-Carlo model predictions of PYTHIA 6.115 (CTEQ3L, \hardcut).  The  four curves correspond to 
the contribution from flavor creation (FIG.~\ref{prd_fig1}), flavor excitation (FIG.~\ref{prd_fig3}),  
shower/fragmentation (FIG.~\ref{prd_fig4}), and the resulting total.}
\label{prd_fig5}
\end{figure}

\begin{figure}[htbp]
\includegraphics[scale=0.5]{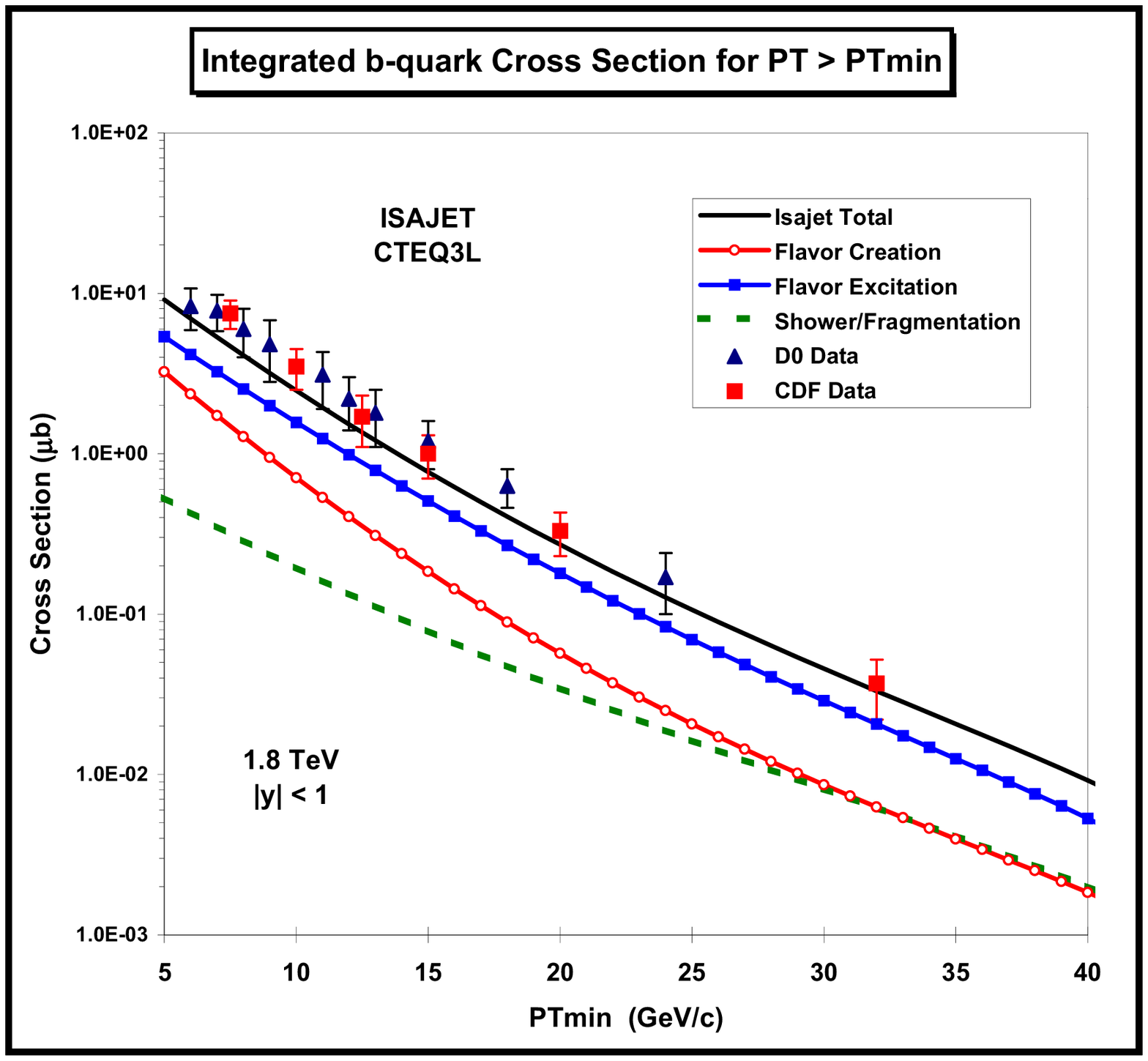}
\caption{Data on the integrated inclusive $b$-quark total cross section  
($p_T>p_T({\rm min})$, \ycut) for proton-antiproton collisions at $1.8\tev$ compared with the 
QCD Monte-Carlo model predictions of ISAJET 7.32 (CTEQ3L, \hardcut).  The  four curves correspond to 
the contribution from flavor creation (FIG.~\ref{prd_fig1}), flavor excitation (FIG.~\ref{prd_fig3}),  
shower/fragmentation (FIG.~\ref{prd_fig4}), and the resulting total.}
\label{prd_fig6}
\end{figure}

\begin{figure}[htbp]
\includegraphics[scale=0.5]{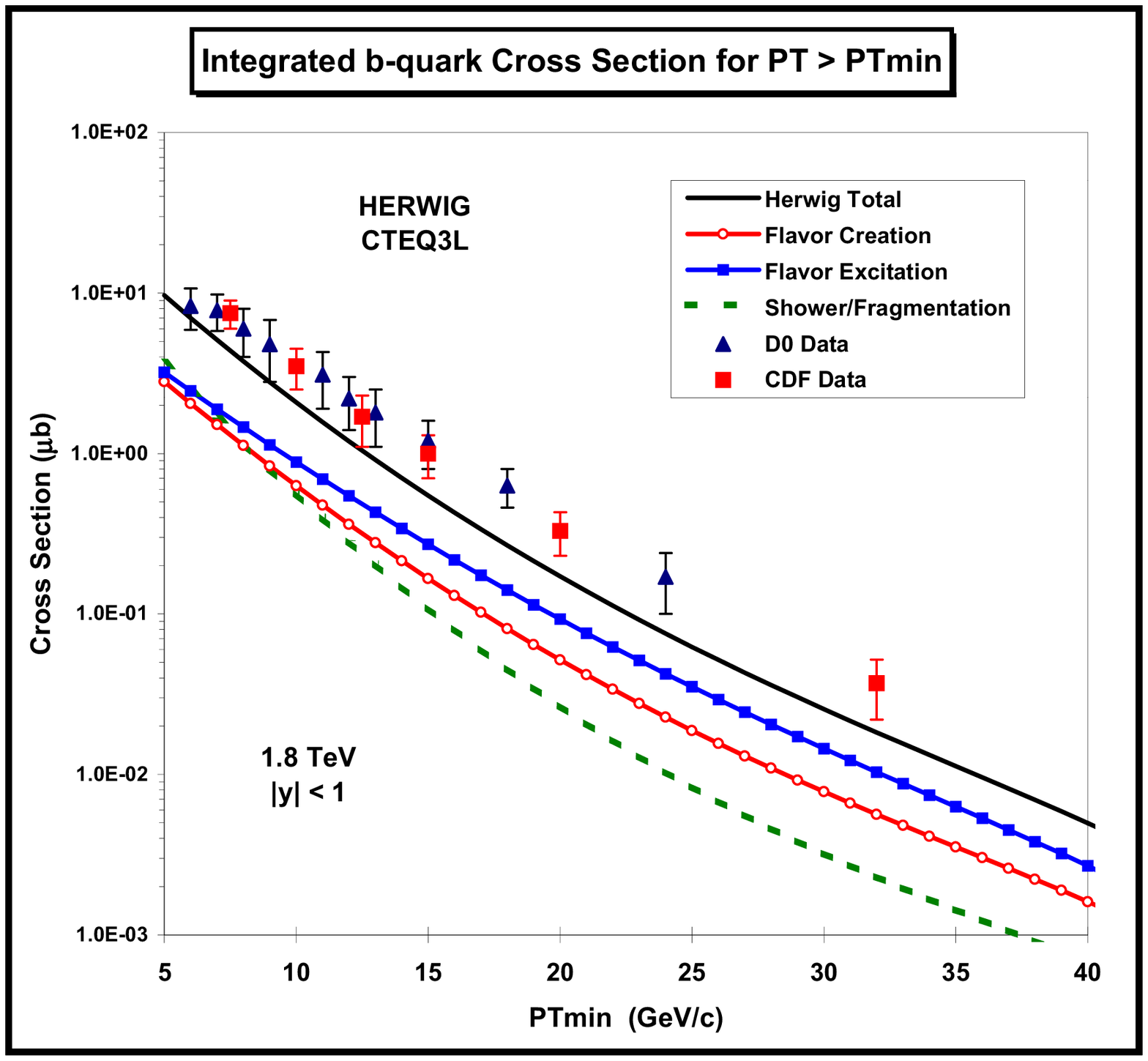}
\caption{Data on the integrated inclusive $b$-quark total cross section  
($p_T>p_T({\rm min})$, \ycut) for proton-antiproton collisions at $1.8\tev$ compared with the 
QCD Monte-Carlo model predictions of HERWIG 5.9 (CTEQ3L, \hardcut).  The  four curves correspond to 
the contribution from flavor creation (FIG.~\ref{prd_fig1}), flavor excitation (FIG.~\ref{prd_fig3}),  
shower/fragmentation (FIG.~\ref{prd_fig4}), and the resulting total.}
\label{prd_fig7}
\end{figure}

\section{Monte-Carlo Generation}

It is not easy to arrive at the QCD Monte-Carlo model predictions for $b$-quark production at the Tevatron.  
Some of the plots presented here contain $8$ million generated events.  In addition, one must handle 
each of the Monte-Carlo models differently in order to get all three contributions; flavor creation, 
flavor excitation, and shower/fragmentation.  

\subsection{Flavor Creation}

The flavor creation contribution to $b$-quark production at the Tevatron is the easiest to generate.  
ISAJET, HERWIG, and PYTHIA all allow the user to select and generate separately the flavor creation 
contribution.  For ISAJET if one runs all QCD and then selects the flavor creation 
terms by monitoring the $2$-to-$2$ subprocess, one gets the same answer as running the flavor creation 
contribution separately.  However, this is not true for HERWIG and PYTHIA.  These two Monte-Carlo models 
only include the $b$-quark mass when one runs the flavor creation terms separately (\ie heavy quark production).  
For HERWIG and PYTHIA, when one runs all QCD the $b$-quark mass is set to zero.  ISAJET always  
includes the $b$-quark mass in the flavor creation subprocesses.  Setting the $b$-quark mass 
equal to zero increases the $b$-quark inclusive flavor creation cross section for \pt $> 5\gevc$ and \ycut\ by roughly 
$40\%$ at $1.8\tev$.   

In summary, one can produce the flavor creation contribution in two ways with ISAJET, either by running 
them separately or by running all QCD.  For HERWIG and PYTHIA one must generate the flavor creation 
terms separately by running the heavy quark option (IPROC = 1705 or MSEL = 5).  For PYTHIA this 
produces only the flavor creation contribution.  However, for HERWIG the heavy quark option produces 
both the flavor creation terms and the flavor excitation terms and one must monitor the $2$-to-$2$ subprocesses 
to separate them.

\subsection{Flavor Excitation}

For ISAJET one can run the flavor excitation contribution separately or one can run all QCD and then select 
out the flavor excitation contribution (one gets the same answer either way).  For HERWIG running 
the heavy quark option gets both the flavor creation and the flavor excitation contributions 
(with the $b$-quark mass included).  For HERWIG, if one runs all QCD and then selects out the flavor 
excitation contribution one gets a different answer since here the $b$ mass is set to zero.  For PYTHIA the 
only way to get the flavor excitation contribution is to run all QCD and then select out the flavor 
excitation contribution, but here one only gets the massless approximation.  Actually, in leading order 
it is not clear how to handle the $b$-quark mass in the flavor excitation subprocesses since one is 
going from an off-shell $b$ (or \bbar) quark in the initial state to an on-shell 
massive $b$ (or \bbar) quark in the final state.  This limits the accuracy of the leading order estimate of 
this contribution.

\subsection{Shower/Fragmentation}
For all three Monte-Carlo models the only way to get the shower/fragmentation contribution 
to $b$-quark production is to run all QCD $2$-to-$2$ subprocesses and then select out the 
shower/fragmentation contribution by monitoring the subprocesses, which requires producing large numbers of 
events.  The shower/fragmentation contribution differs greatly among the Monte-Carlo models for two reasons. The 
first is due to different fragmentation schemes.  ISAJET uses independent fragmentation, 
while HERWIG and PYTHIA do not.  The second difference arises from the way the QCD Monte-Carlo 
produce parton showers.  ISAJET uses a leading-log picture in which the partons within the shower are 
ordered according to their invariant mass.  Kinematics requires that the invariant mass of daughter 
partons are less than the invariant mass of the parent.  HERWIG and PYTHIA modify the leading-log 
picture to include ``color coherence effects" which leads to ``angle ordering" within the parton shower.

The flavor excitation and shower/fragmentation $2$-to-$2$ subprocesses diverge as the parton transverse 
momentum, \pthard, goes to zero.  To get a finite result one must regulate the divergences or specify a minimum 
value for \pthard.  For HERWIG and ISAJET we take \hardcut.  PYTHIA regulates the perturbative $2$-to-$2$ parton 
scattering differential cross sections, which allow one to take  \pthard $> 0$.  For PYTHIA $b$-quark predictions 
with \pt($b$-quark) $> 5\gevc$ are independent of the whether one takes \hardcut\ or \pthard $> 0$.  

\begin{figure}[htbp]
\includegraphics[scale=0.5]{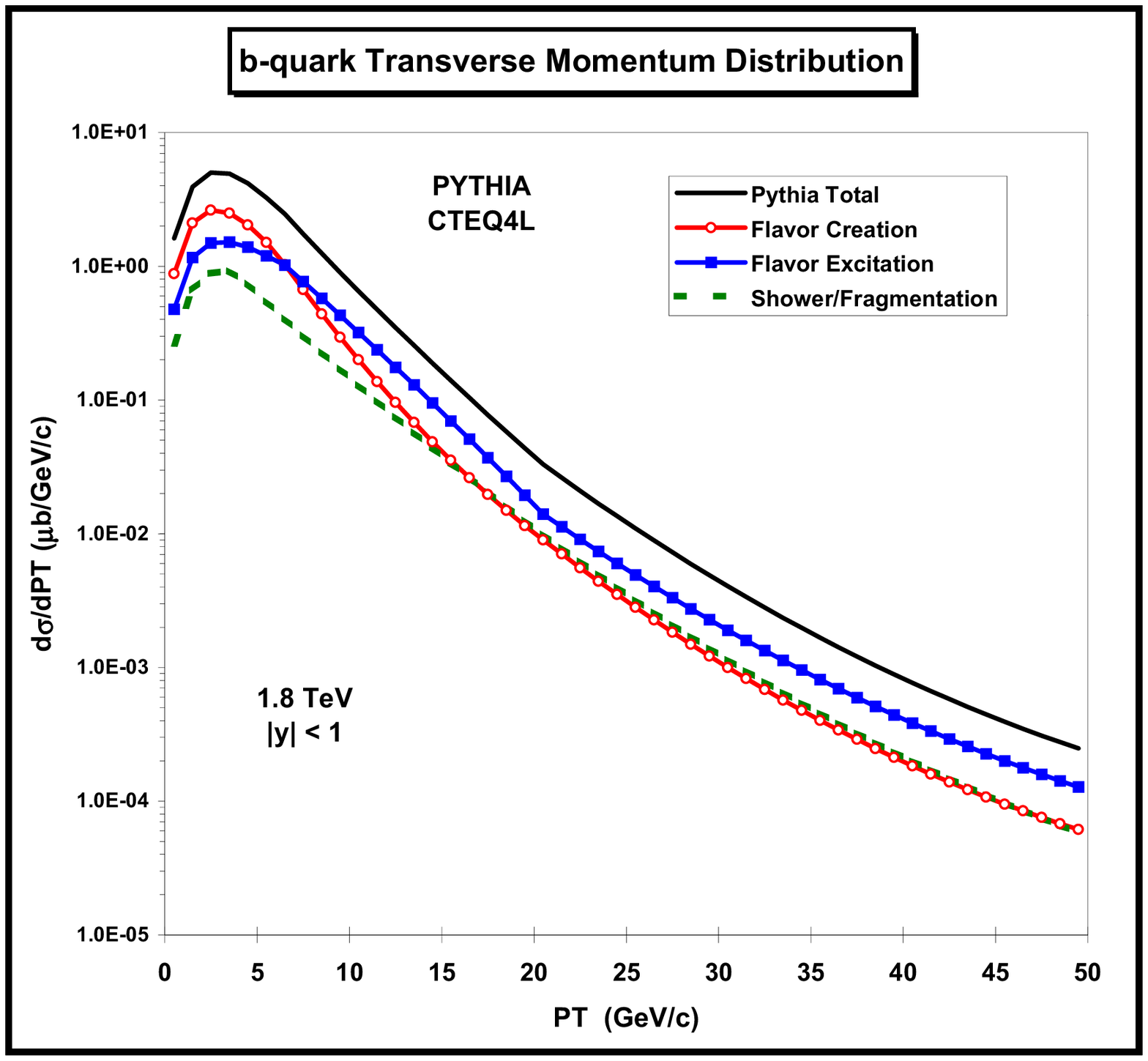}
\caption{The inclusive $b$-quark differential cross section, $d\sigma/dp_T$,  for \ycut\ 
for proton-antiproton collisions at $1.8\tev$ resulting from PYTHIA 6.158 (CTEQ4L, \pthard $> 0$).  The  four curves correspond to 
the contribution from flavor creation (FIG.~\ref{prd_fig1}), flavor excitation (FIG.~\ref{prd_fig3}),  
shower/fragmentation (FIG.~\ref{prd_fig4}), and the resulting total.}
\label{prd_fig8}
\end{figure}

\begin{figure}[htbp]
\includegraphics[scale=0.5]{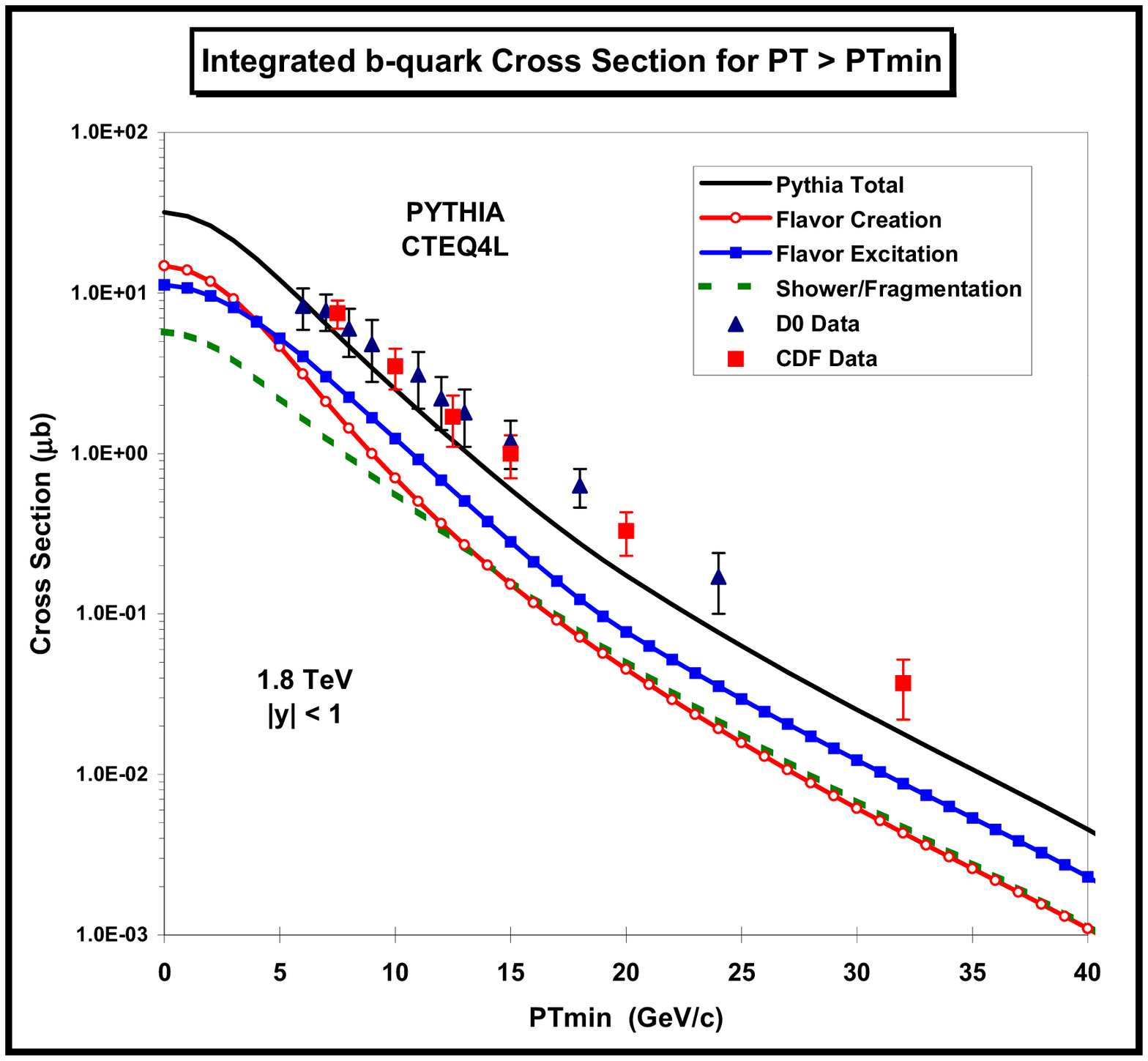}
\caption{Data on the integrated inclusive $b$-quark total cross section  
($p_T>p_T({\rm min})$, \ycut) for proton-antiproton collisions at $1.8\tev$ compared with the 
QCD Monte-Carlo model predictions of PYTHIA 6.158 (CTEQ4L, \pthard $> 0$).  
The  four curves correspond to 
the contribution from flavor creation (FIG.~\ref{prd_fig1}), flavor excitation (FIG.~\ref{prd_fig3}),  
shower/fragmentation (FIG.~\ref{prd_fig4}), and the resulting total.}
\label{prd_fig9}
\end{figure}

\section{The Inclusive Cross-Section}

The data from CDF and D0 \cite{CDF1,CDF2,D0,wester} on the integrated inclusive $b$-quark total cross-section are compared with 
the QCD Monte-Carlo model predictions in FIGS.~\ref{prd_fig5}-~\ref{prd_fig7}, with the CTEQ3L 
structure functions and \hardcut.  The  four curves in each of the plots correspond to the 
contribution to $b$-quark production from  flavor creation, flavor excitation,  shower/fragmentation, 
and the resulting overall total.  After adding the contributions from all three sources 
PHYTHIA (CTEQ3L) agrees fairly well with the data.  ISAJET (CTEQ3L) is a bit below the data because of a 
smaller shower/fragmentation contribution.  HERWIG (CTEQ3L) is also slightly below the data due to a
smaller flavor excitation component.  One does not expect precise agreement from these leading-log estimates.  
However, the qualitative agreement indicates that probably nothing unusual is happening 
in $b$-quark production at the Tevatron.

FIG.~\ref{prd_fig8} and FIG.~\ref{prd_fig9} show the differential inclusive $b$-quark \pt\ cross section 
and the integrated total cross section, respectively, from PYTHIA (CTEQ4L), with \pthard $> 0$.  PYTHIA regulates 
the divergences in the perturbative $2$-to-$2$ parton scattering differential cross sections and these plots 
show a smooth behavior down to \pt($b$-quark) equal to zero.

\begin{figure}[htbp]
\includegraphics[scale=0.6]{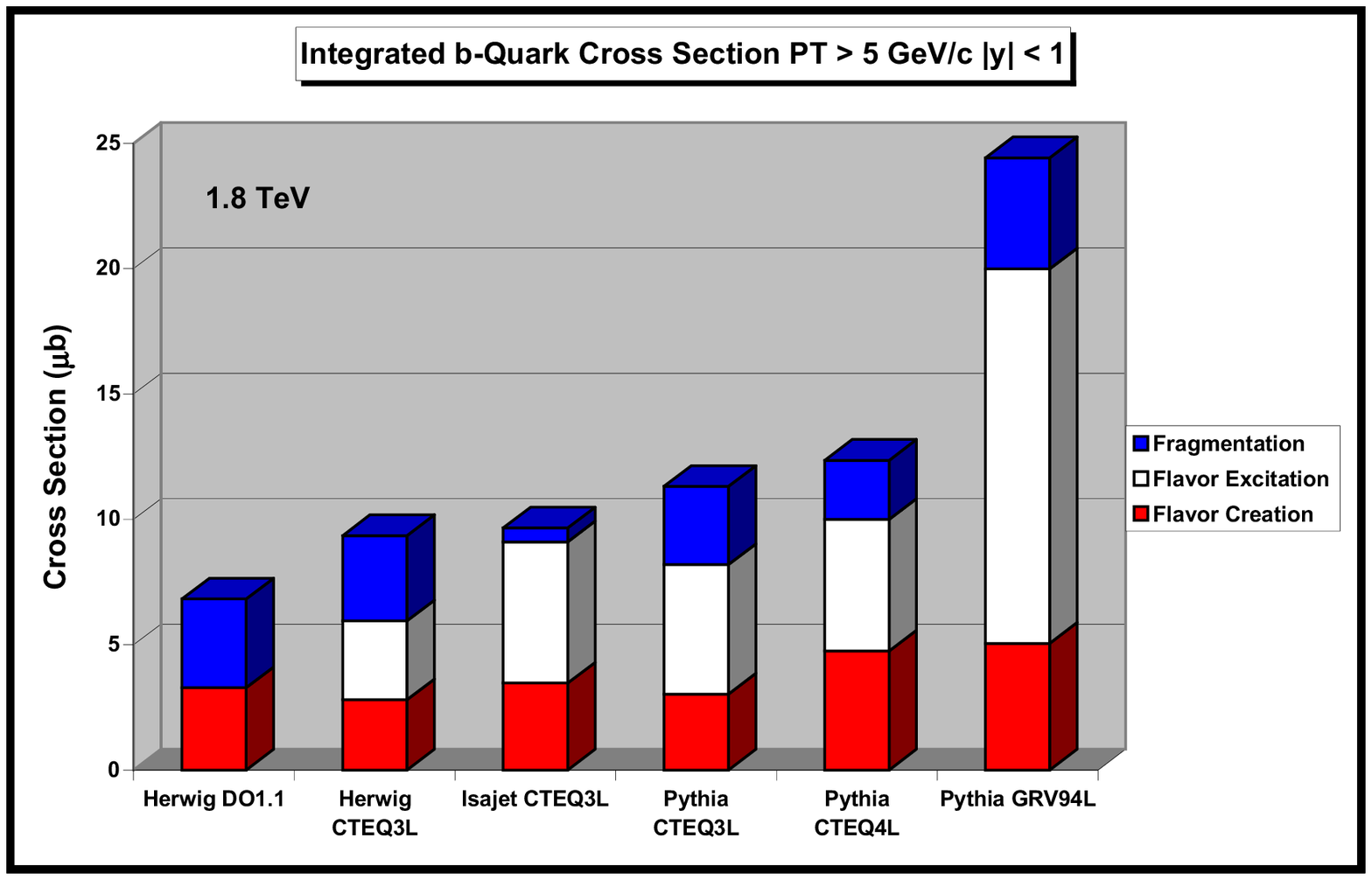}
\caption{Predictions of HERWIG 5.9 (DO1.1), HERWIG 5.9 (CTEQ3L), ISAJET 7.32 (CTEQ3L), 
PYTHIA 6.115 (CTEQ3L), PYTHIA 6.158 (CTEQ4L), and PYTHIA 6.115 (GRV94L) for the integrated 
inclusive $b$-quark total cross section 
($p_T>5\gevc$, \ycut) for proton-antiproton collisions at $1.8\tev$.
The contributions from flavor creation (FIG.~\ref{prd_fig1}), flavor excitation (FIG.~\ref{prd_fig3}),  
shower/fragmentation (FIG.~\ref{prd_fig4}) are shown together with the resulting sum (overall height of box).}
\label{prd_fig10}
\end{figure}

FIG.~\ref{prd_fig10} shows the predictions of HERWIG (DO1.1), HERWIG (CTEQ3L), ISAJET (CTEQ3L), 
PYTHIA (CTEQ3L), PYTHIA (CTEQ4L), and PYTHIA (GRV94L) for the integrated $b$-quark total cross section  
(\pt$>5\gevc$, \ycut) at $1.8\tev$. The contributions from flavor creation, flavor excitation, and  
shower/fragmentation are shown together with the overall resulting sum (overall height of box).  
The DO1.1 structure functions include only four flavors and the probability of 
finding a $b$ (or \bbar) quark within the proton is set identically equal to zero.  Thus, running with 
the DO1.1 structure functions results in no flavor excitation.  The differences among the Monte-Carlo models
for flavor excitation are due to the different ways the models handle the $b$-quark mass in this 
subprocess.  

The Monte-Carlo model predictions for the shower/fragmentation contribution differ considerably.  This is not 
surprising since ISAJET uses independent fragmentation, while HERWIG and PYTHIA do not; and HERWIG and PYTHIA 
modify the leading-log picture of parton showers to include ``color coherence effects", while ISAJET does not.  

The flavor creation contribution to inclusive $b$-quark production (\pt$>5\gevc$, \ycut) is predicted by 
HERWIG (CTEQ3L) and PYTHIA (CTEQ3L) to be about $25\%$ of the overall $b$-quark production rate from 
all three sources at $1.8\tev$, while for ISAJET (CTEQ3L) it is about $35\%$ of the total.

\begin{figure}[htbp]
\includegraphics[scale=0.6]{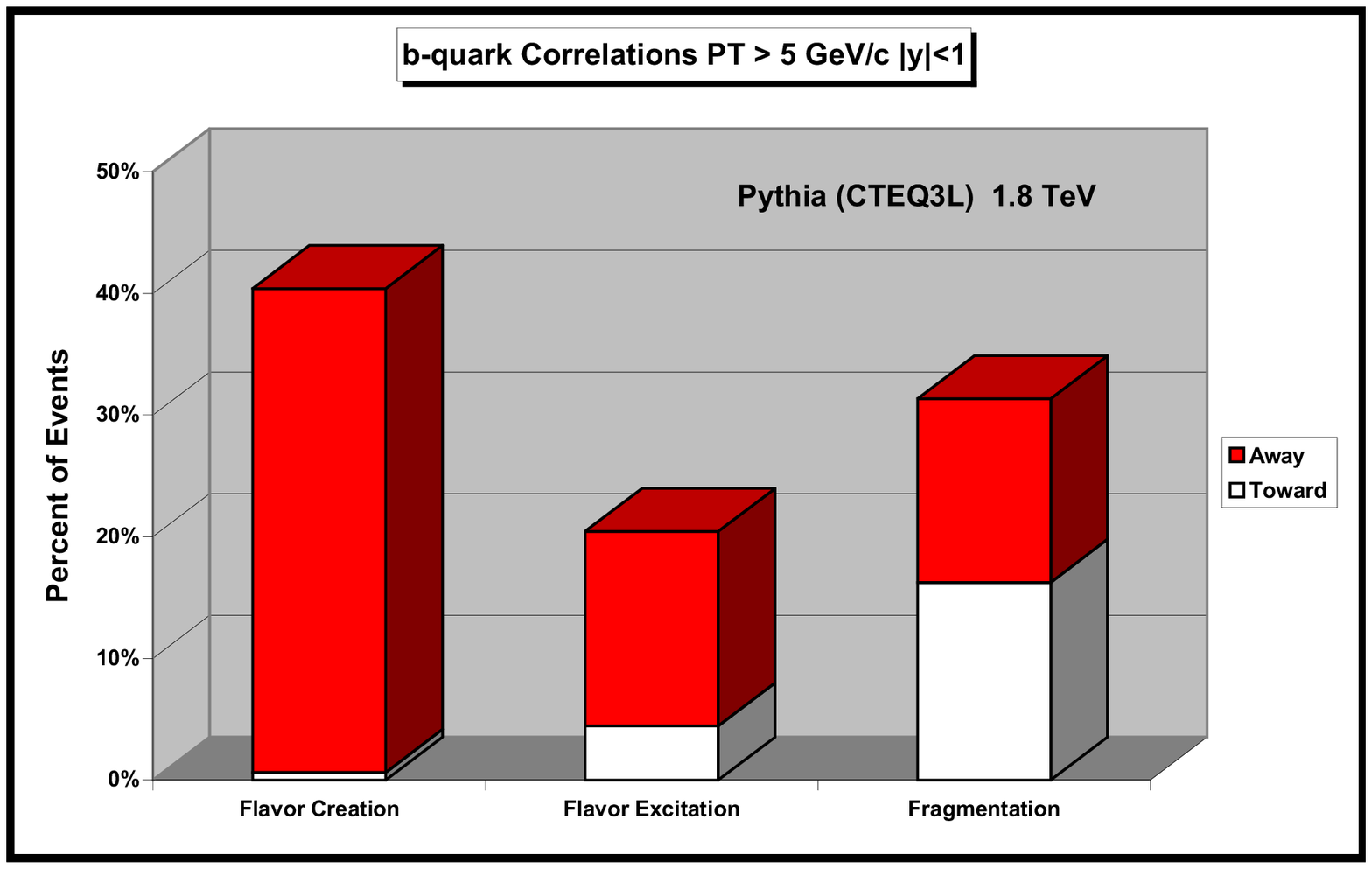}
\caption{Predictions of  PYTHIA 6.115 (CTEQ3L, \hardcut) for the probability of finding a \bbar-quark 
with $p_T > 5\gevc$ and \ycut\ in an event with a $b$-quark with $p_T > 5\gevc$ and \ycut\ for proton-antiproton 
collisions at $1.8\tev$.  The contribution from the ``toward" (\delphi$<90^\circ$) and the ``away" (\delphi$>90^\circ$) 
region of the $b$-quark are shown for 
flavor creation (FIG.~\ref{prd_fig1}), flavor excitation (FIG.~\ref{prd_fig3}), and
shower/fragmentation (FIG.~\ref{prd_fig4}), where \delphi\ is the azimuthal angle between the $b$ and \bbar-quark.}
\label{prd_fig11}
\end{figure}

\begin{figure}[htbp]
\includegraphics[scale=0.6]{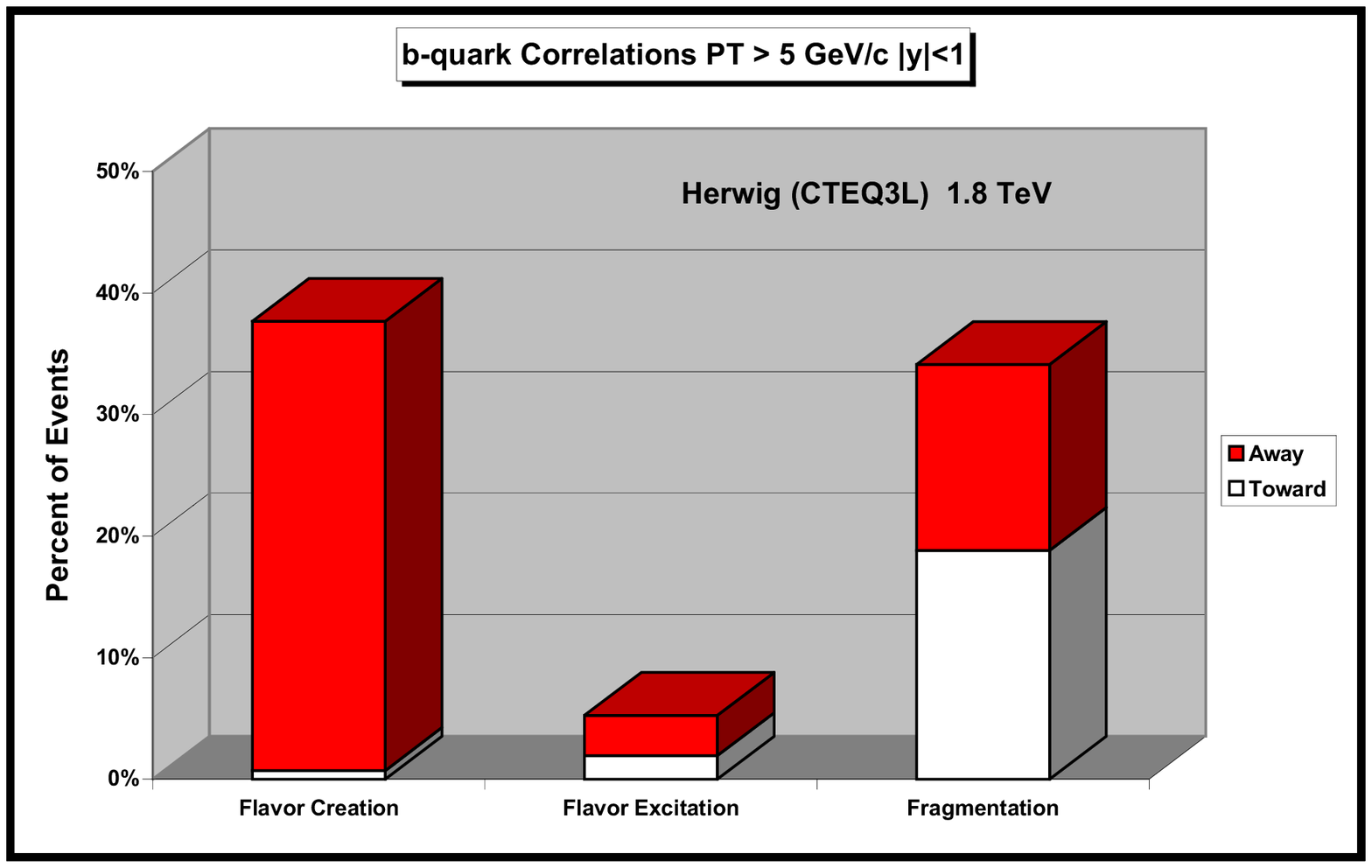}
\caption{Predictions of  HERWIG 5.9 (CTEQ3L, \hardcut) for the probability of finding a \bbar-quark 
with $p_T > 5\gevc$ and \ycut\ in an event with a $b$-quark with $p_T > 5\gevc$ and \ycut\ for proton-antiproton 
collisions at $1.8\tev$.  The contribution from the ``toward" (\delphi$<90^\circ$) and the ``away" (\delphi$>90^\circ$) 
region of the $b$-quark are shown for 
flavor creation (FIG.~\ref{prd_fig1}), flavor excitation (FIG.~\ref{prd_fig3}), and
shower/fragmentation (FIG.~\ref{prd_fig4}), where \delphi\ is the azimuthal angle between the $b$ and \bbar-quark.}
\label{prd_fig12}
\end{figure}

\begin{figure}[htbp]
\includegraphics[scale=0.6]{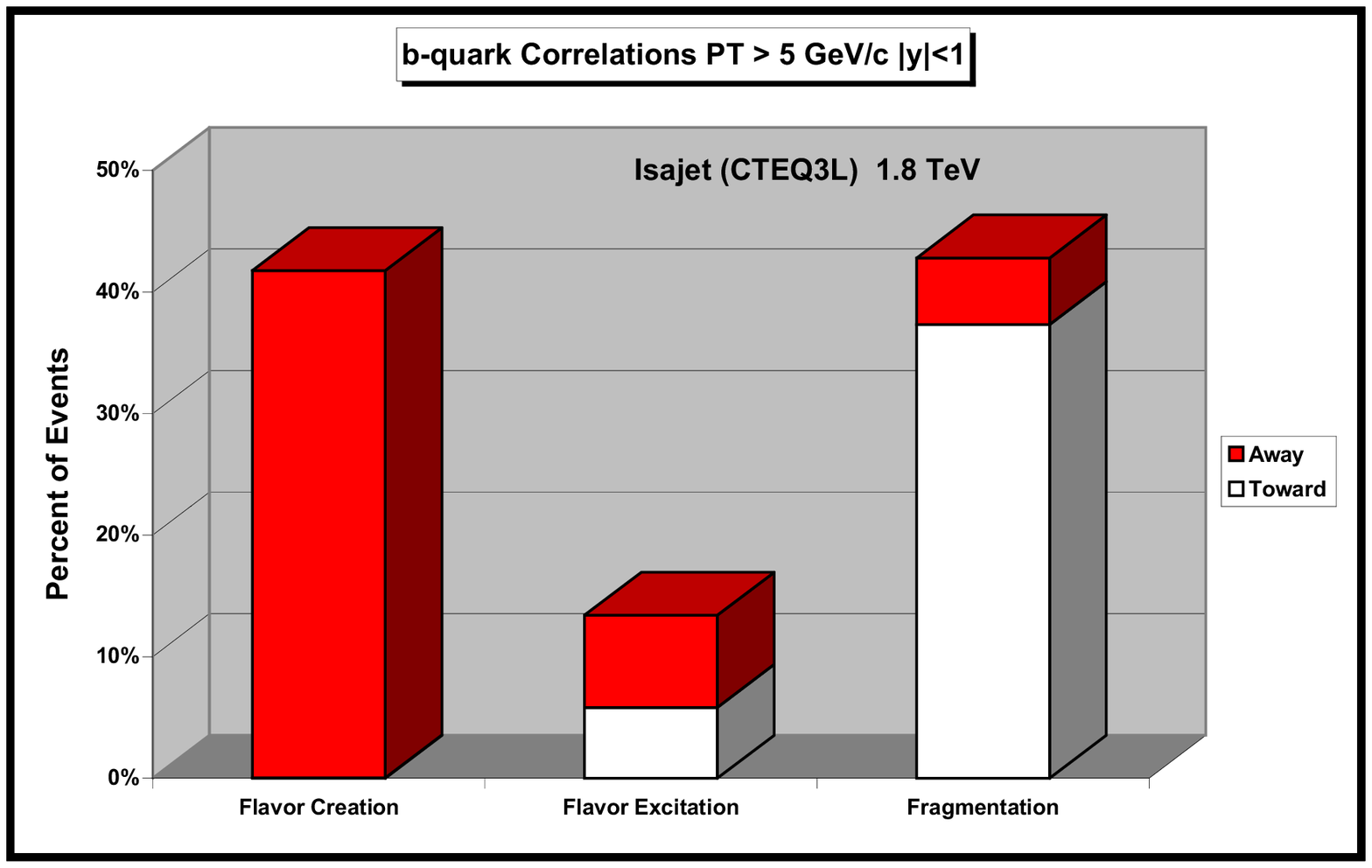}
\caption{Predictions of  ISAJET 7.32 (CTEQ3L, \hardcut) for the probability of finding a \bbar-quark 
with $p_T > 5\gevc$ and \ycut\ in an event with a $b$-quark with $p_T > 5\gevc$ and \ycut\ for proton-antiproton 
collisions at $1.8\tev$.  The contribution from the ``toward" (\delphi$<90^\circ$) and the ``away" (\delphi$>90^\circ$) 
region of the $b$-quark are shown for 
flavor creation (FIG.~\ref{prd_fig1}), flavor excitation (FIG.~\ref{prd_fig3}), and
shower/fragmentation (FIG.~\ref{prd_fig4}), where \delphi\ is the azimuthal angle between the $b$ and \bbar-quark.}
\label{prd_fig13}
\end{figure}

\section{Correlations}

Clearly the three sources of $b$-quarks; flavor creation, flavor excitation, and shower/fragmentation have 
quite different topological structures and the correlations between the $b$ and \bbar\ quarks are quite different.  

\subsection{Simple ``Toward" and ``Away" Probabilities}

FIGS.~\ref{prd_fig11} -~\ref{prd_fig13} show the QCD Monte-Carlo model predictions for some simple correlations.  
Here one requires a $b$-quark to be in the region \pt $> 5\gevc$ and \ycut\ and then asks for the probability 
of finding a  \bbar-quark in the same region, \pt $> 5\gevc$ and \ycut.  Furthermore, one breaks this 
probability into two terms, ``toward" and ``away".  The ``toward" region corresponds to \delphi $< 90^\circ$ and 
the ``away" region has \delphi $> 90^\circ$, where $\Delta\phi$ is the azimuthal angle between the $b$ and  \bbar\ quark.  
The models predict that for flavor creation the probability that both the $b$ and  \bbar\ quark lie in 
the region \pt $> 5\gevc$ and \ycut\ is about $40\%$, with the  \bbar-quark almost always on the ``away-side" 
of the $b$-quark.

The QCD Monte-Carlo models differ considerably on the correlations for the shower/fragmentation contribution.  
For this contribution all three models predict that both the $b$ and  \bbar\ quark lie in the 
region \pt $> 5\gevc$ and \ycut\ around $30\%$-$40\%$ of the time, which is comparable to the flavor 
creation contribution.  However, ISAJET predicts that for the shower/fragmentation contribution that 
the  \bbar-quark is almost always on the ``toward-side" of the $b$-quark, while HERWIG and PYTHIA predict 
about equal amounts of ``toward" and ``away" for this contribution. 

As one would expect, all the models predict that it is not very likely to find both the $b$ and the  \bbar-quark in the 
region \pt $> 5\gevc$ and \ycut\ for the flavor excitation contribution.  For flavor excitation 
either the $b$-quark or the  \bbar-quark is part of the ``underlying" event (see FIG.~\ref{prd_fig3}).  
The models differ greatly on the correlations for the flavor excitation contribution.  For this contribution 
PYTHIA predicts that both the $b$ and  \bbar-quark lie in the region \pt $> 5\gevc$ and \ycut\ around $20\%$ 
of the time, while ISAJET predicts $13\%$ and HERWIG gives $5\%$ for this probability.  All three models predict 
that for flavor excitation when both the $b$ and \bbar-quark are found in the region \pt $> 5\gevc$ and \ycut\   
that the  \bbar-quark is about equally likely to be on ``toward" or ``away" side of the $b$-quark.

\begin{figure}[htbp]
\includegraphics[scale=0.6]{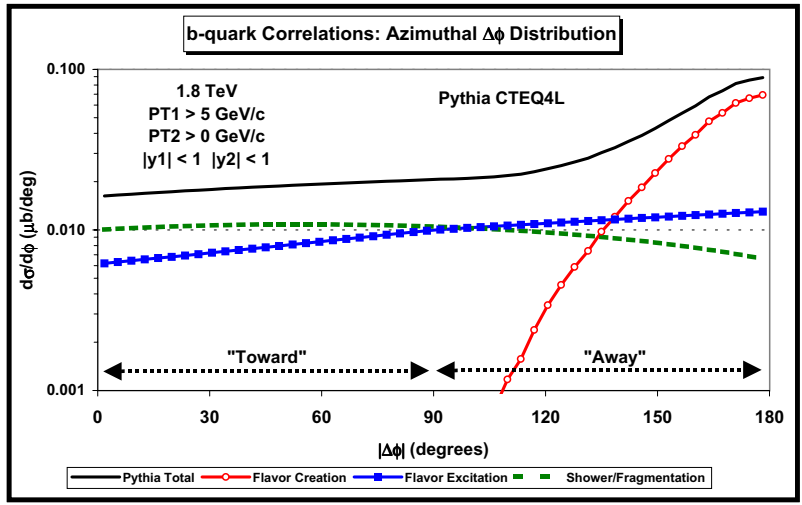}
\caption{Predictions of  PYTHIA 6.158 (CTEQ4L, \pthard$>0$) for the azimuthal angle, $\Delta\phi$, between a 
$b$-quark with $p_{T1}>5\gevc$ and  $|y_1|<1$ and a  \bbar-quark with $p_{T2}>0$ and $|y_2|<1$ in 
proton-antiproton collisions at $1.8\tev$.  The curves correspond to $d\sigma/d\Delta\phi$ ($\mu$b/$^\circ$) for 
flavor creation (FIG.~\ref{prd_fig1}), flavor excitation (FIG.~\ref{prd_fig3}),
shower/fragmentation (FIG.~\ref{prd_fig4}, and the resulting total. 
The ``toward" (\delphi$<90^\circ$) and the ``away" (\delphi$>90^\circ$) region of the $b$-quark are labeled. 
(\textit{Note the logarithmetic scale.})}
 \label{prd_fig14}
\end{figure}

\begin{figure}[htbp]
\includegraphics[scale=0.6]{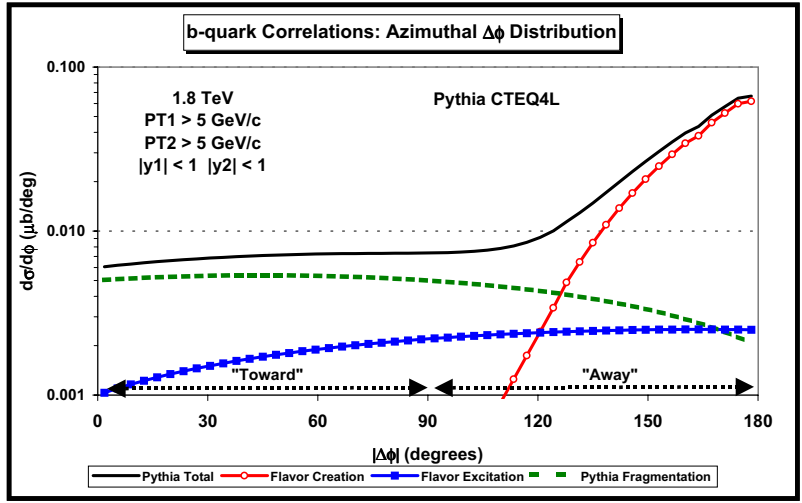}
\caption{Predictions of  PYTHIA 6.158 (CTEQ4L, \pthard$>0$) for the azimuthal angle, $\Delta\phi$, between a 
$b$-quark with $p_{T1}>5\gevc$ and $|y_1|<1$ and a  \bbar-quark with $p_{T2}>5\gevc$ and $|y_2|<1$ in 
proton-antiproton collisions at $1.8\tev$.  The curves correspond to $d\sigma/d\Delta\phi$ ($\mu$b/$^\circ$) for 
flavor creation (FIG.~\ref{prd_fig1}), flavor excitation (FIG.~\ref{prd_fig3}),
shower/fragmentation (FIG.~\ref{prd_fig4}), and the resulting total. 
The ``toward" (\delphi$<90^\circ$) and the ``away" (\delphi$>90^\circ$) region of the $b$-quark are labeled. 
(\textit{Note the logarithmetic scale.})}
 \label{prd_fig15}
\end{figure}

\subsection{$\Delta\phi$ Distribution}

FIG.~\ref{prd_fig14} shows the predictions of PYTHIA for the azimuthal angle, $\Delta\phi$, between a $b$-quark 
with  $p_{T1}> 5\gevc$ and $|y_1|<1$ and  \bbar\ quark with $p_{T2}> 0$ and $|y_2|<1$.
FIG.~\ref{prd_fig15} is the same as FIG.~\ref{prd_fig14} except both the $b$ and \bbar\ quark are 
required to have $p_T > 5\gevc$.  Notice that the relative amounts of flavor excitation and shower/fragmentation  
decrease compared to flavor creation when one demands that both the $b$ and \bbar\ quarks to have $p_T > 5\gevc$.
Integrating the curves in FIG.~\ref{prd_fig15} over the ``toward" and ``away" 
regions correspond to the simple probabilities discussed in Section IV(a).  Clearly, the ``toward" region is very sensitive 
to the presence of the flavor excitation and shower/fragmentation terms.

\begin{figure}[htbp]
\includegraphics[scale=0.6]{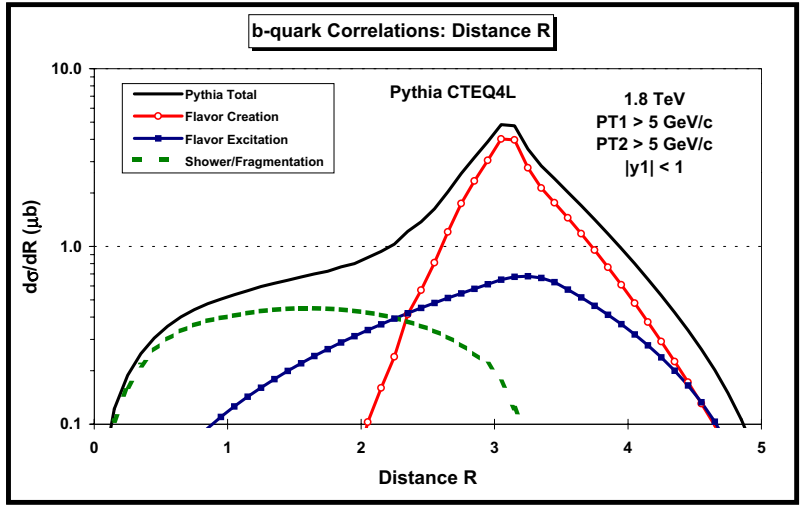}
\caption{Predictions of PYTHIA 6.158 (CTEQ4L, \pthard$>0$) for the distance, $R$, in \etaphi\ space 
between a $b$-quark with $p_{T1}>5\gevc$ and $|y_1|<1$  and a \bbar-quark with $p_{T2}>5\gevc$ proton-antiproton 
collisions at $1.8\tev$.  The curves correspond to $d\sigma/dR$ ($\mu$b) for 
flavor creation (FIG.~\ref{prd_fig1}), flavor excitation (FIG.~\ref{prd_fig3}),
shower/fragmentation (FIG.~\ref{prd_fig4}), and the resulting total. 
(\textit{Note the logarithmetic scale.})}
 \label{prd_fig16}
\end{figure}

\subsection{Distance ``R"}

FIG.~\ref{prd_fig16} shows the predictions of PYTHIA for the 
distance, $R$, in \etaphi\ space between a $b$-quark with $p_{T1}> 5\gevc$ and $|y_1|<1$ and  \bbar-quark 
with $p_{T2}> 5\gevc$, where $R=\sqrt{(\eta_1-\eta_2)^2+(\phi_1-\phi_2)^2}$ and $\eta$ is the pseudo-rapidity.  
One can see that the region $R < 1$ isolates the shower/fragmentation contribution.

\subsection{Rapidity Correlations}

\begin{figure}[htbp]
\includegraphics[scale=0.6]{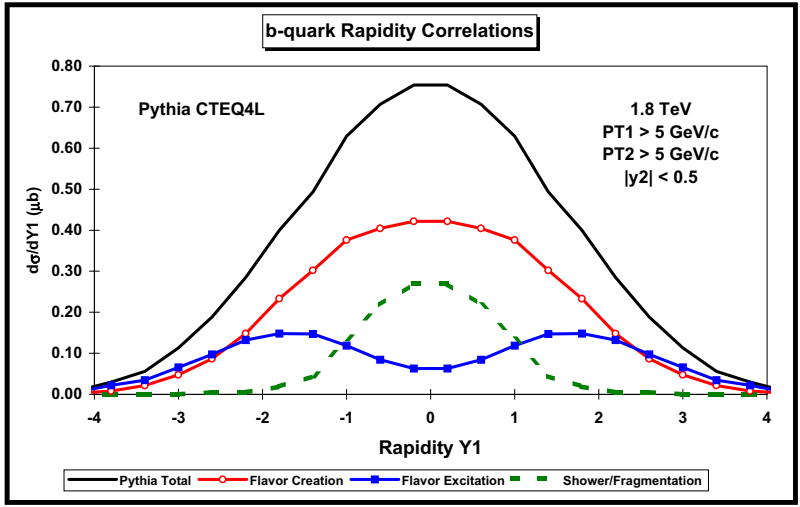}
\caption{Predictions of PYTHIA 6.158 (CTEQ4L, \pthard$>0$) for the rapidity, $y_1$, of a $b$-quark 
with $p_{T1}>5\gevc$ for events with a  \bbar-quark with $p_{T2}>5\gevc$ and $|y_2|<0.5$ 
in proton-antiproton collisions at $1.8\tev$.  The curves correspond to $d\sigma/dy_1$ ($\mu$b) 
for flavor creation (FIG.~\ref{prd_fig1}), flavor excitation (FIG.~\ref{prd_fig3}),
shower/fragmentation (FIG.~\ref{prd_fig4}), and the resulting total.}
 \label{prd_fig17}
\end{figure}

FIG.~\ref{prd_fig17} shows the predictions of PYTHIA for the rapidity, $y_1$ of a $b$-quark with 
$p_{T1}> 5\gevc$ for events with a  \bbar-quark with $p_{T2}> 5\gevc$ and $|y_2|<0.5$. Here the 
flavor excitation contribution behaves much differently than the other two terms. However, 
the resulting sum of the three contributions looks similar to the flavor creation term.

\begin{figure}[htbp]
\includegraphics[scale=0.6]{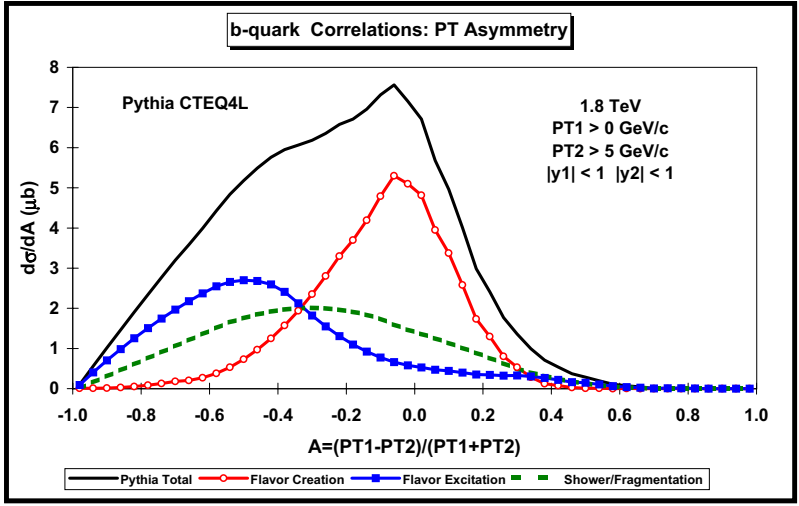}
\caption{Prediction of PYTHIA 6.158 (CTEQ4L, \pthard$>0$) for the \pt\ asymmetry,  $A=(p_{T1}-p_{T2})/(p_{T1}+p_{T2})$, 
for events with a $b$-quark with $p_{T1}>0$ and  $|y_1| < 1.0$ and a  \bbar-quark 
with $p_{T2}>5\gevc$ and $|y_2|<1.0$ in proton-antiproton collisions at $1.8\tev$.
The curves correspond to $d\sigma/dA$ ($\mu$b) 
for flavor creation (FIG.~\ref{prd_fig1}), flavor excitation (FIG.~\ref{prd_fig3}),
shower/fragmentation (FIG.~\ref{prd_fig4}), and the resulting total.}
 \label{prd_fig18}
\end{figure}

\subsection{The Transverse Momentum Asymmetry}

FIG.~\ref{prd_fig18} shows the predictions of PYTHIA for the \pt\ asymmetry, $A=(p_{T1}-p_{T2})/(p_{T1}+p_{T2})$, 
for events with a $b$-quark with $p_{T1}> 0$ and $|y_1|<1$ and a  \bbar-quark with
$p_{T2}> 5\gevc$ and $|y_2|<1$.  If there were no initial-state radiation (see FIG.~\ref{prd_fig1}) and  
no ``intrinsic" transverse momentum, the flavor creation terms would produce zero asymmetry.  Also, as one 
would expect (FIG.~\ref{prd_fig3}), the flavor excitation terms have, on the average,  a large 
transverse momentum asymmetry. Requiring \pt\ asymmetries with a magnitude greater than around $60\%$ isolates the flavor 
excitation contribution.

\begin{figure}[htbp]
\includegraphics[scale=0.5]{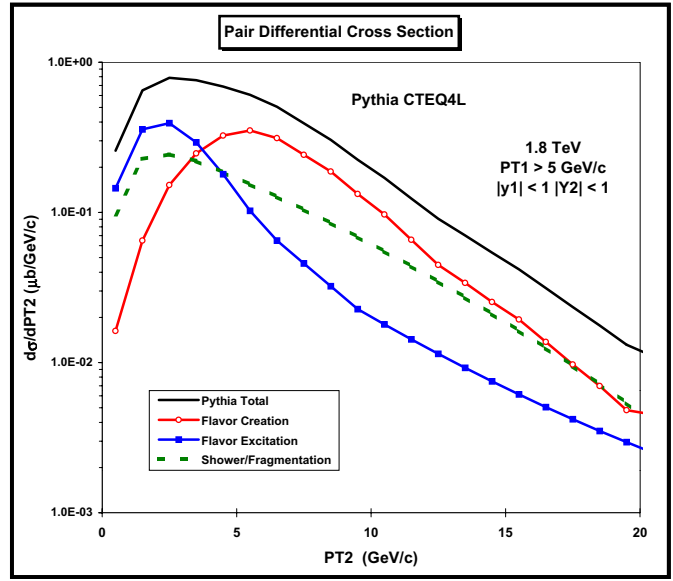}
\caption{Prediction of PYTHIA 6.158 (CTEQ4L, \pthard$>0$) for the transverse momentum, $p_{T2}$, of a  
\bbar-quark with $|y_2|<1.0$ for events with a $b$-quark with $p_{T1}>5\gevc$ and $|y_1|<1.0$ in proton-antiproton 
collisions at $1.8\tev$.  The curves correspond to $d\sigma/dp_{T2}$ ($\mu$b/GeV/c) for 
flavor creation (FIG.~\ref{prd_fig1}), flavor excitation (FIG.~\ref{prd_fig3}),
shower/fragmentation (FIG.~\ref{prd_fig4}), and the resulting total.
(\textit{Note the logarithmetic scale.})}
 \label{prd_fig19}
\end{figure}

\begin{figure}[htbp]
\includegraphics[scale=0.5]{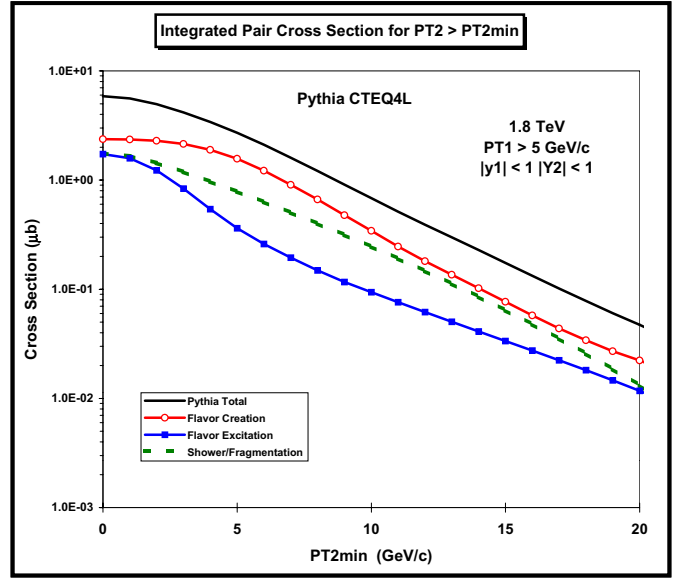}
\caption{
Prediction of  PYTHIA 6.158 (CTEQ4L, \pthard$>0$) for the integrated pair cross section for a  
\bbar-quark with $p_{T2}>p_{T2}({\rm min})$ and $|y_2|<1.0$ and
a $b$-quark with $p_{T1}>5\gevc$ and $|y_1|<1.0$ in proton-antiproton 
collisions at $1.8\tev$.  The curves correspond to $\sigma$ ($\mu$b) for 
flavor creation (FIG.~\ref{prd_fig1}), flavor excitation (FIG.~\ref{prd_fig3}),
shower/fragmentation (FIG.~\ref{prd_fig4}), and the resulting total.
(\textit{Note the logarithmetic scale.})}
 \label{prd_fig20}
\end{figure}

\subsection{Double Differential Cross Section}

FIG.~\ref{prd_fig19} shows the predictions of PYTHIA for the transverse momentum, $p_{T2}$, of a  \bbar-quark 
with $|y_2|<1$ for events with a b-quark with $p_{T1}> 5\gevc$ and $|y_1|<1$.  As the transverse momentum 
asymmetry in FIG.~\ref{prd_fig18} indicated, the flavor creation contribution peaks at 
$p_{T2}\approx 5\gevc$ (\ie small \pt\ asymmetry), while the flavor excitation and shower/fragmentation contributions 
peak at low \pt\ (\ie larger \pt\ asymmetry).

FIG.~\ref{prd_fig20} shows the predictions of PYTHIA for the integrated  \bpair\ pair cross section for 
a  \bbar-quark with $p_{T2} > p_{T2}({\rm min})$ and $|y_2|<1$ and b-quark 
with $p_{T1}> 5\gevc$ and $|y_1|<1$.  Notice how the three terms contribute in different proportions 
to the \bpair\ pair cross section in FIG.~\ref{prd_fig20} and the inclusive single $b$-quark cross section 
in FIG.~\ref{prd_fig9}.  Flavor excitation is a major contributor to the single inclusive cross section, 
while flavor creation dominates the pair cross section.

\section{Summary and Conclusions}

The leading-log order QCD hard scattering Monte-Carlo models of HERWIG, ISAJET, and PYTHIA have been used to 
study the sources of $b$-quarks at the Tevatron.  The reactions responsible of producing $b$-quarks are 
separated into flavor creation, flavor excitation, and shower/fragmentation.

\vskip 0.1in
\centerline{\it Flavor Creation}\nopagebreak

The production of a  \bpair\ pair via gluon fusion or annihilation of light quarks is easy to generate 
and all three QCD Monte-Carlo models (\textit{with the same structure functions}) predict roughly the same
cross section and similar \bpair\ correlations from these terms.  At the Tevatron all three Monte-Carlo 
models predict that the flavor creation contribution to $b$-quark production is less than $35\%$ of the 
overall inclusive b-quark production rate (\pt$>5\gevc$, \ycut) from all three sources.

\vskip 0.1in
\centerline{\it Flavor Excitation}\nopagebreak

The source of $b$-quarks resulting from the scattering of a $b$-quark or \bbar-quark out of the 
initial-state into the final-state by a gluon or by a light quark or antiquark is difficult 
to generate and depends sensitively on the parton distribution functions.  The QCD Monte-Carlo model 
predictions differ somewhat, however, it seems likely that at the Tevatron the flavor 
excitation contribution to the inclusive $b$-quark cross section is comparable to or greater than the 
contribution from flavor creation.  The \bpair\ correlations resulting from the flavor excitation 
term are much different than those arising from flavor creation. In particular, the distribution 
in the azimuthal angle,  $\Delta\phi$, between the $b$ and the \bbar-quark peaks sharply in 
the ``backward" (\delphi $> 90^\circ$) region for flavor creation, while flavor excitation produces 
an approximately flat \delphi\ distribution.  Also, the flavor excitation terms have, on the average, a large 
transverse momentum asymmetry between the $b$ and \bbar-quark, while the flavor creation terms 
produce a \pt\ asymmetry near zero.

\vskip 0.1in
\centerline{\it Shower/Fragmentation}\nopagebreak

The production of \bpair\ pairs within a parton-shower or during the fragmentation process of gluons and light quarks 
and antiquarks is an important source at the Tevatron.  The QCD Monte-Carlo models predictions differ considerably 
for this contribution.  However, at the Tevatron the shower/fragmentation contribution to the 
inclusive b-quark cross section might be comparable to the contribution from flavor creation.  This source can 
be isolated by looking for \bpair\ pairs with $R < 1$, where $R$ is the distance in \etaphi\ space between 
the $b$ and \bbar-quark.

\vskip 0.2in
The QCD Monte-Carlo model leading-log estimates of the flavor excitation and the shower/fragmentation 
contributions to $b$-quark production are uncertain and should not be taken too seriously.  
However, it seems likely that at the Tevatron all three sources of $b$-quarks, flavor creation, flavor excitation, 
and shower/fragmentation are important \cite{TS}. One does not expect precise agreement from leading-log estimates.  
On the other hand, the qualitative agreement shown indicates that probably 
nothing unusual is happening in $b$-quark production at the Tevatron.  Furthermore, in Run II at the Tevatron we 
should be able to isolate experimentally the individual contributions to $b$-quark production by studying  
\bpair\ correlations in detail.

\begin{acknowledgments}
I would like to thank the CDF B Group for several stimulating discussions.  Also, I would like to thank Henry Frisch, 
Michelangelo Mangano, and Torbj\"orn Sj\"ostrand for helpful remarks.
\end{acknowledgments}


\end{document}